\documentclass[reviewcopy]{elsart}
\usepackage{amssymb}
\usepackage{graphicx}
\usepackage[letterpaper]{geometry}

\begin{document}

\newcommand{\barl}{\bar{\lambda}}

\newcommand{\barp}{\bar{p}}
\begin{flushright}
hep-ph/0603037\\
ANL-HEP-PR-06-30
\end{flushright}

\begin{frontmatter}

\title{All-orders Resummation for \\ 
Diphoton Production at Hadron Colliders}

\author{Csaba Bal\'{a}zs, Edmond L. Berger, Pavel Nadolsky}

\address{High Energy Physics Division, Argonne National Laboratory,\\
9700 Cass Ave., Argonne, IL 60439, USA }

\author{C.-P. Yuan}

\address{Department of Physics and Astronomy, Michigan State University, \\
 East Lansing, MI 48824, USA}

\begin{abstract}
\begin{minipage}[c]{6.5in}
We present a QCD calculation of the transverse momentum distribution 
of photon pairs produced at hadron colliders, including all-orders 
soft-gluon resummation valid at next-to-next-to-leading
logarithmic accuracy.  We specify the region of phase space in 
which the calculation is most reliable, compare our results with 
data from the Fermilab Tevatron, and make predictions for the 
Large Hadron Collider.  The uncertainty of predictions for production 
of diphotons from fragmentation of final-state quarks is examined.   
\end{minipage}
\end{abstract}
\date{March 3, 2006}
\end{frontmatter}

{\em {Introduction.}} 
A Higgs boson with mass between 115 and 140 GeV may be identified at 
hadron colliders through its decay into a pair of energetic photons, 
a challenging prospect at the Large Hadron Collider (LHC) in view of the 
intense background from hadronic production of non-resonant photon 
pairs~\cite{LHCexpts:1999fr}. 
Theoretical predictions of these background processes may be of 
substantial value in aiding search strategies. Moreover, the 
perturbative quantum chromodynamics (QCD) calculation of photon-pair 
production is of theoretical interest in its own right, and data from the 
Tevatron collider offer an opportunity to compare and test results against  
experiment. 

In this paper, we present a new calculation of the diphoton cross 
section in perturbative QCD. We include contributions from all 
perturbative subprocesses (quark-antiquark, quark-gluon, antiquark-gluon, 
and gluon-gluon) to next-to-leading (NLO) accuracy.  In 
addition, to describe properly the behavior of the transverse momentum 
$Q_T$ distribution of the pairs in the region in which $Q_T < Q$, where 
$Q$ is the invariant mass of the photon pair, we include the all-orders 
resummation of soft and collinear logarithmic contributions up to 
next-to-next-to-leading log (NNLL) accuracy.  This calculation goes 
beyond the previous resummation treatments of diphoton 
production~\cite{Balazs:1997hv,Balazs:1999yf}. Its components are 
summarized briefly below, and a more complete discussion is presented
elsewhere~\cite{DiphotonLongPRD}.

A full treatment of photon pair production requires that we address the 
contributions from non-perturbative processes, such as $\pi$ and $\eta$ 
meson decays,  
and the quasi-collinear fragmentation of quarks and gluons into photons.  
Elaborate isolation procedures are applied by the experiments to reduce 
these long-distance contributions, procedures that are only approximately 
reproducible theoretically.   
Some final-state fragmentation contributions invariably survive the
isolation, especially at the LHC, where the efficiency of isolation
is reduced by event pile-up and the large number of energetic hadronic
fragments in each event. A new feature of diphoton production, with respect 
to single photon production, is the prospect that both photons may be 
produced from fragmentation of the same final-state parton.  This 
fragmentation contribution is expected to be most influential 
in the region in which both the diphoton invariant mass and 
the separation $\Delta \varphi$ between the azimuthal angles of the two photons 
are relatively small, $Q < Q_T$ and $\Delta \varphi < \pi/4$.  

Diphoton production is characterized by large
radiative corrections, distributed in a complex pattern over the accessible
phase space. The influence of initial-state gluon radiation on the 
predicted transverse momentum distributions can be evaluated 
to all orders with the Collins-Soper-Sterman (CSS) resummation 
procedure~\cite{Collins:1984kg}, the method that we follow.  
Our results are implemented in a Monte-Carlo integration program RESBOS.  
We use a simple, efficient approximation for the fragmentation contributions.
We compare our results with data from the Collider
Detector at Fermilab (CDF) collaboration at $p\bar{p}$ collision
energy $\sqrt{S}=1.96$ TeV and integrated 
luminosity 207 $\mbox{pb}^{-1}$~\cite{Acosta:2004sn}, and we observe 
good agreement.  
We make several suggestions for a further more differential analysis of 
the data that would allow refined tests of our calculation.  In view 
of theoretical uncertainties associated with the fragmentation component 
of the cross section, and the presence of other large radiative corrections, 
we question the conclusion in Ref.~\cite{Acosta:2004sn}
that the inclusion of single-photon fragmentation contributions within 
the NLO calculation of Ref.~\cite{Binoth:1999qq} uniquely explains the
observed kinematic distributions of the diphotons at the Tevatron.
We also include predictions for diphoton production at the LHC.  

{\em Analytical Calculation.}
The CSS resummation method is used in Refs.~\cite{Balazs:1997hv,Balazs:1999yf} 
to treat the direct production of photon pairs from $q\bar{q}$,
$q^{^{\!\!\!\!\!\!\!(-)}}g,$ and $gg$ scattering. The NLO perturbative
cross sections (i.e., cross sections of ${\mathcal{O}}(\alpha_{s})$
in the $q\bar{q}$ and $qg$ channels~\cite{Aurenche:1985yk,Berger:1990et,Bailey:1992br},
and ${\mathcal{O}}(\alpha_{s}^{3})$ in the $gg$ 
channel~\cite{Balazs:1999yf,deFlorian:1999tp,Bern:2001df,Bern:2002jx}) are 
included
as a part of the resummed cross section. Singular logarithms arise
in the NLO cross sections when the transverse momentum $Q_{T}$ of
the $\gamma\gamma$ pair is much smaller than its invariant mass $Q$.
These logarithms are resummed into a Sudakov exponent (composed of
two anomalous-dimension functions $A(\mu)$ and $B(\mu)$) and convolutions
of the conventional parton densities $f_{a}(x,\mu_{F})$ with Wilson
coefficient functions $C$. In Refs.~\cite{Balazs:1997hv,Balazs:1999yf},
the functions $A(\mu)$, $B(\mu)$ and $C$ are evaluated up to order
$\alpha_{s}^{2}$, $\alpha_{s},$ and $\alpha_{s}$, respectively.
An approximate expression is used there for the $C$-function of order 
$\alpha_{s}$
in the $gg$ subprocess (borrowed from the $gg\rightarrow\mbox{Higgs}$
resummed cross section). In this work, we include the exact
$C$-function of order $\alpha_{s}$ for $gg\rightarrow\gamma\gamma X$
\cite{Nadolsky:2002gj} and ${\mathcal{O}}(\alpha_{s}^{2})$ expressions
for $A(\mu)$ and $B(\mu)$ in all 
subprocesses~\cite{Nadolsky:2002gj,deFlorian:2000pr,deFlorian:2001zd}.
These enhancements elevate the accuracy of the resummed prediction
to the NNLL level. We use an improved model for the non-perturbative 
contributions at large impact parameter~\cite{Konychev:2005iy}. 
When expanded in a series in $\alpha_{s}$,
the resummed predictions for the total rate, $\gamma \gamma$ invariant 
mass, and $\gamma \gamma$ rapidity ($y$) distributions are equal 
to the fixed-order QCD cross 
sections, augmented by higher-order contributions from the integrated 
$Q_{T}$ logs. The resummed $Q_T$ distribution is well-behaved as 
$Q_T \rightarrow 0$, unlike its fixed-order counterpart which is 
singular in this limit.  As $Q_T$ grows, our resummed cross section 
crosses the perturbative NLO cross section at $Q_{T}\sim Q$, and, 
for each $Q$ and $y$, we switch from the resummed to the NLO cross 
section for values of $Q_T$ above this point.  

A fragmentation singularity arises in the matrix element when the 
momentum of a photon is collinear with that of an outgoing quark or gluon. 
The fragmentation singularities do not appear in the resummed terms since 
those correspond to initial-state radiation.  At
the lowest order, the fragmentation singularity appears in the 
$qg\rightarrow\gamma\gamma q$ channel and is proportional to 
$P_{\gamma\leftarrow q}(z)/(n-4)$ in $n$-dimensional regularization, 
where $P_{\gamma\leftarrow q}(z)$ is the $q \rightarrow \gamma$ splitting 
function, and $z$ is the fraction of the fragmenting quark's light-cone 
momentum carried
by the photon.  The fragmentation singularity
is subtracted from the direct contribution. It is resummed in the
photon fragmentation function $D_{\gamma}(z)$ through the introduction  
of a {}``one-fragmentation'' contribution 
$q+g\rightarrow(q\stackrel{frag}{\longrightarrow\gamma} X)+\gamma$,
where $``(\stackrel{frag}{q\longrightarrow\gamma X})$'' denotes collinear
production of a photon from a quark. For a wide class of two-to-two 
partonic processes, such as $q \bar q \rightarrow q \bar q$, etc., there is 
a second type of {}``one-fragmentation'' 
contribution that arises in low-mass photon-pair 
production ($Q < Q_T$).  In this case, a final-state parton may fragment 
into a low-mass pair of photons, a process described by a 
different fragmentation function $D_{\gamma \gamma}(z_1, z_2)$.  This new 
contribution is not included in the existing calculations.  
{}``Two-fragmentation'' contributions arise in processes like 
$g+g\rightarrow(q\stackrel{frag}{\longrightarrow}\gamma X)+(\bar{q}\stackrel{frag}{\longrightarrow}\gamma X)$
and involve convolutions with two functions $D_{\gamma}(z)$ (one
per photon).  

Isolation constraints must be imposed on the inclusive photon
cross sections before the comparison with data.  Isolation can
be applied to the cross sections at each order of 
$\alpha_{s}$~\cite{Berger:1995cc,Catani:1998yh,Catani:2002ny}.
The magnitude of the fragmentation contribution is controlled by the
isolation procedure chosen and can be strongly affected by tuning
the quasi-experimental isolation model. An isolation condition in a 
typical measurement requires the hadronic activity to be
minimal (e.g., comparable to the underlying event) in the immediate
neighborhood of each candidate photon. Candidate photons may be rejected
because of energy deposit nearby in the hadronic calorimeter, which
introduces dependence on the calorimeter cell geometry, or because 
hadronic tracks are present near the photons. A theory calculation
may approximate the experimental isolation by requiring the full energy
of the hadronic remnants to be less than a threshold {}``isolation energy''
$E_{T}^{iso}$ in the cone $\Delta R=\sqrt{\Delta\eta^{2}+\Delta\varphi^{2}}$
around each photon, with $\Delta\eta$ and $\Delta\varphi$ being
the separations of the hadronic remnant(s) from the photon in the
plane of pseudo-rapidity $\eta$ and azimuthal angle $\varphi$. The
two photons must also be separated in the $\eta-\varphi$ plane by
an amount exceeding the approximate angular size $\Delta R_{\gamma\gamma}$
of one calorimeter cell. The values of $E_{T}^{iso},$ $\Delta R$,
and $\Delta R_{\gamma\gamma}$ serve as crude characteristics of the
actual measurement. The size of the fragmentation contributions depends 
tangibly on the assumed values of 
$E_{T}^{iso},$ $\Delta R$, and $\Delta R_{\gamma\gamma}$,
as is shown below. 

We find it sufficient in our work to use a simplified fragmentation  
model to represent the isolated cross section.  
We regularize the fragmentation region by imposing a combination of
a sharp cutoff $E_{T}^{iso}$ on the transverse energy $E_{T}$ of
the final-state quark or gluon and smooth cone isolation~\cite{Frixione:1998jh}.
We impose quasi-experimental isolation by rejecting an event if (a)
the separation $\Delta r=\sqrt{(\eta-\eta_{\gamma})^{2}+(\varphi-\varphi_{\gamma})^{2}}$
between the final-state parton and one of the photons is less than
$\Delta R$, and (b) $E_{T}$ of the parton is larger than $E_{T}^{iso}$.
This condition excludes the singular fragmentation contributions 
in the finite-order $qg$ cross section at $\Delta r<\Delta R$ and 
$E_{T}>E_{T}^{iso}$. The 
fragmentation contributions at $\Delta r<\Delta R$
and $E_{T}<E_{T}^{iso}$ are suppressed by rejecting events in the
$\Delta R$ cone that satisfy $E_{T}<\chi(\Delta r)$, where $\chi(\Delta r)$
is a smooth function satisfying $\chi(0)=0,$ $\chi(\Delta R)=E_{T}^{iso}$.
This {}``smooth cone isolation''~\cite{Frixione:1998jh} transforms
the non-integrable fragmentation singularity associated with $D_{\gamma}(z)$
into an integrable singularity of a magnitude dependent on the functional
form of $\chi(\Delta r)$. Infrared safety of the cross sections is
preserved as a result of smoothness of $\chi(\Delta r)$. The cross
section for direct contributions is rendered finite by this prescription
without the explicit introduction of fragmentation functions $D_{\gamma}(z)$.
For our smooth function, we choose 
$\chi(\Delta r)=E_{T}^{iso}(1-\cos\Delta r)^{2}/(1-\cos\Delta R)^{2}$. 
Modifications to the function $\chi(\Delta r)$ lead to only mild variations 
of our predicted $Q_{T}$ distribution for $Q_{T}<E_{T}^{iso}$.

In our calculation, we use the electroweak parameters
\cite{Eidelman:2004wy} $G_{F}=1.16639\times10^{-5}~{\textrm{GeV}}^{-2},$
$m_{Z}=91.1882~{\textrm{GeV}},$ and $m_{W}=80.419~{\textrm{GeV}}$.   
We use two-loop expressions for the running electromagnetic
and strong couplings $\alpha(\mu)$ and $\alpha_{S}(\mu)$, as well
as the NLO parton distribution function set CTEQ6M~\cite{Pumplin:2002vw} 
and set 1 of the NLO photon fragmentation functions from 
Ref.~\cite{Bourhis:1997yu}.
Our choices of the renormalization and factorization scales are the same 
as in Ref.~\cite{Balazs:1997hv}; in particular, we set $\mu_{R}=\mu_{F}=Q$ 
in the finite-order perturbative expressions.

In impact parameter ($b$) space, used in the CSS resummation procedure, we must 
integrate into the non-perturbative region of large $b$. Contributions from 
this region are known to be suppressed at high energies~\cite{Berger:2002ut}, but 
it is important nevertheless to evaluate the expected residual uncertainties. We 
use a model for the non-perturbative contributions
({}``revised $b_{*}$ model'') based
on the analysis of Drell-Yan pair and $Z$ boson production in 
Ref.~\cite{Konychev:2005iy}.  A non-perturbative Sudakov function for the factorization 
constant $C_{3}=2e^{-\gamma_{E}}\approx1.123$
is used here to describe the non-perturbative terms in the 
leading $q\bar{q}\rightarrow\gamma\gamma$ channel~\cite{Konychev:2005iy}. 
We neglect possible corrections to the non-perturbative contributions
arising from the final-state soft radiation in the $qg$ channel,
as well as additional $\sqrt{S}$ dependence affecting Drell-Yan-like
processes at $x\lesssim10^{-2}$ \cite{Berge:2004nt}, as those exceed the 
accuracy of the present measurements at the Tevatron. The non-perturbative
function in the $gg\rightarrow\gamma\gamma$ channel is approximated
by multiplying the non-perturbative function for the $q\bar{q}$ channel
by the ratio $C_{A}/C_{F}=9/4$ of the color factors $C_{A}=3$ and
$C_{F}=4/3$ for the leading soft contributions in the $gg$ and $q\bar{q}$
channels. Comparing our results based on the {}``revised $b_{*}$ model'' 
with those obtained with the original $b_{*}$ approach, we find at most 
10\% differences in our predicted $d\sigma/dQ_T$ at the lowest values of 
$Q_T$ at the Tevatron collider energy, and smaller differences at larger 
values of $Q_T$, all well within the experimental uncertainties. The 
differences are even smaller at the LHC energy~\cite{DiphotonLongPRD}. 

{\em Comparison with Tevatron Data.} Our analysis provides a 
calculation of the triple-differential cross section 
$d \sigma/dQ dQ_T d \Delta \varphi$.  Its relevance is especially pertinent 
for the transverse momentum $Q_T$ distribution in the region 
$Q_T \le Q$, for fixed values of diphoton mass $Q$.  It would 
be best to compare our multi-differential 
distribution with experiment, but the published collider data 
tend to be presented in the form of single-differential 
distributions in $Q$, $Q_T$, and $\Delta \varphi$, after integration 
over the other variables. We follow suit in order to 
make comparisons with the Tevatron collider data, but we comment on 
the features that can be explored if more differential studies are 
made.  In accord with CDF, we impose the 
cuts $|y_{\gamma}|<0.9$ 
on the rapidity of each photon, and 
$p_{T}^{\gamma}>p_{Tmin}^{\gamma}= $~14 (13) GeV on the 
transverse momentum of the harder (softer) photon in each 
$\gamma\gamma$ pair. We choose $E_{T}^{iso}=1$ GeV, 
$\Delta R=0.4,$ and $\Delta R_{\gamma\gamma}=0.3$, unless stated otherwise.

\begin{figure}
\begin{center}\includegraphics[%
  width=0.5\textwidth,
  keepaspectratio]{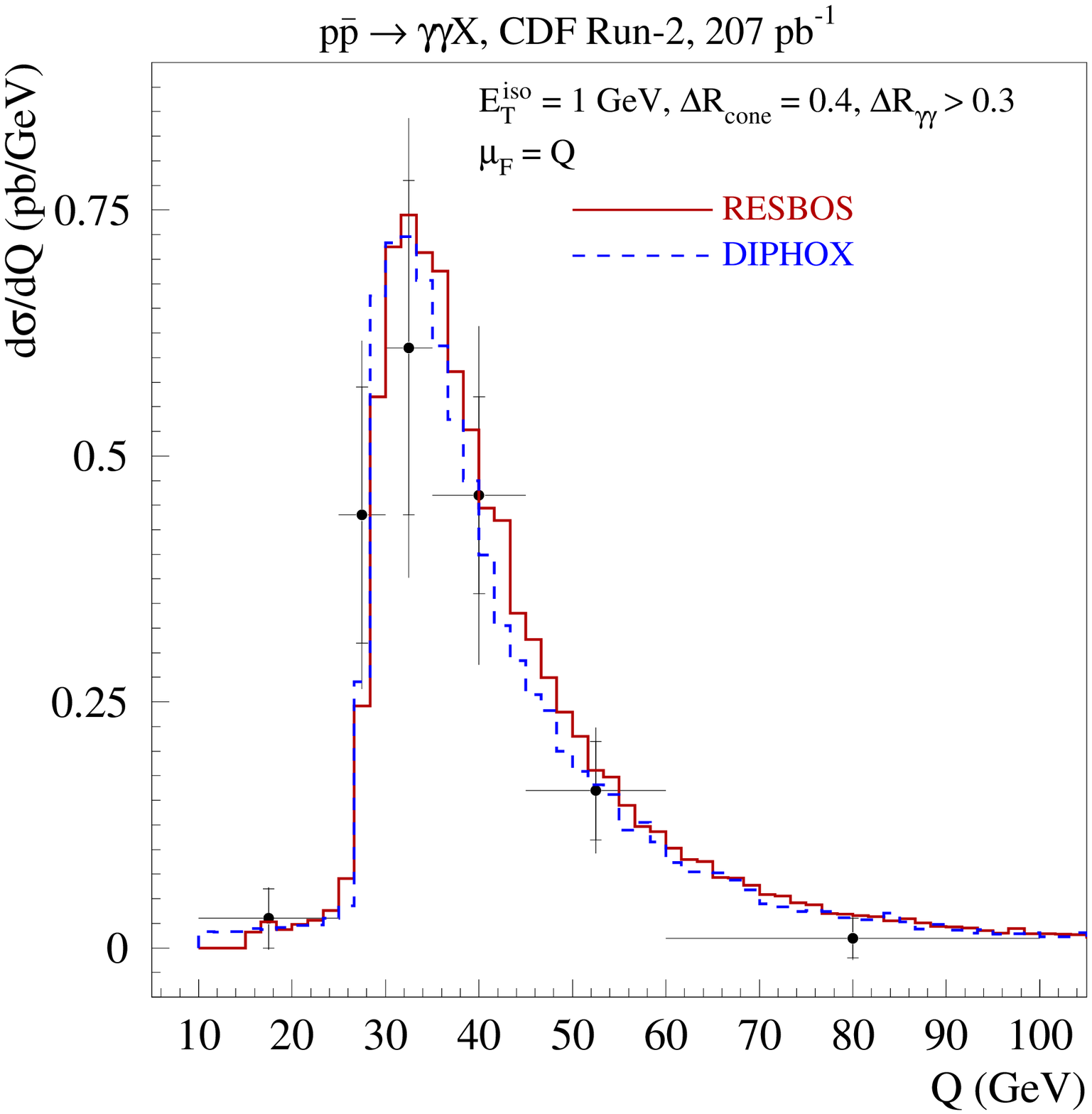}~~\includegraphics[%
  width=0.5\textwidth,
  keepaspectratio]{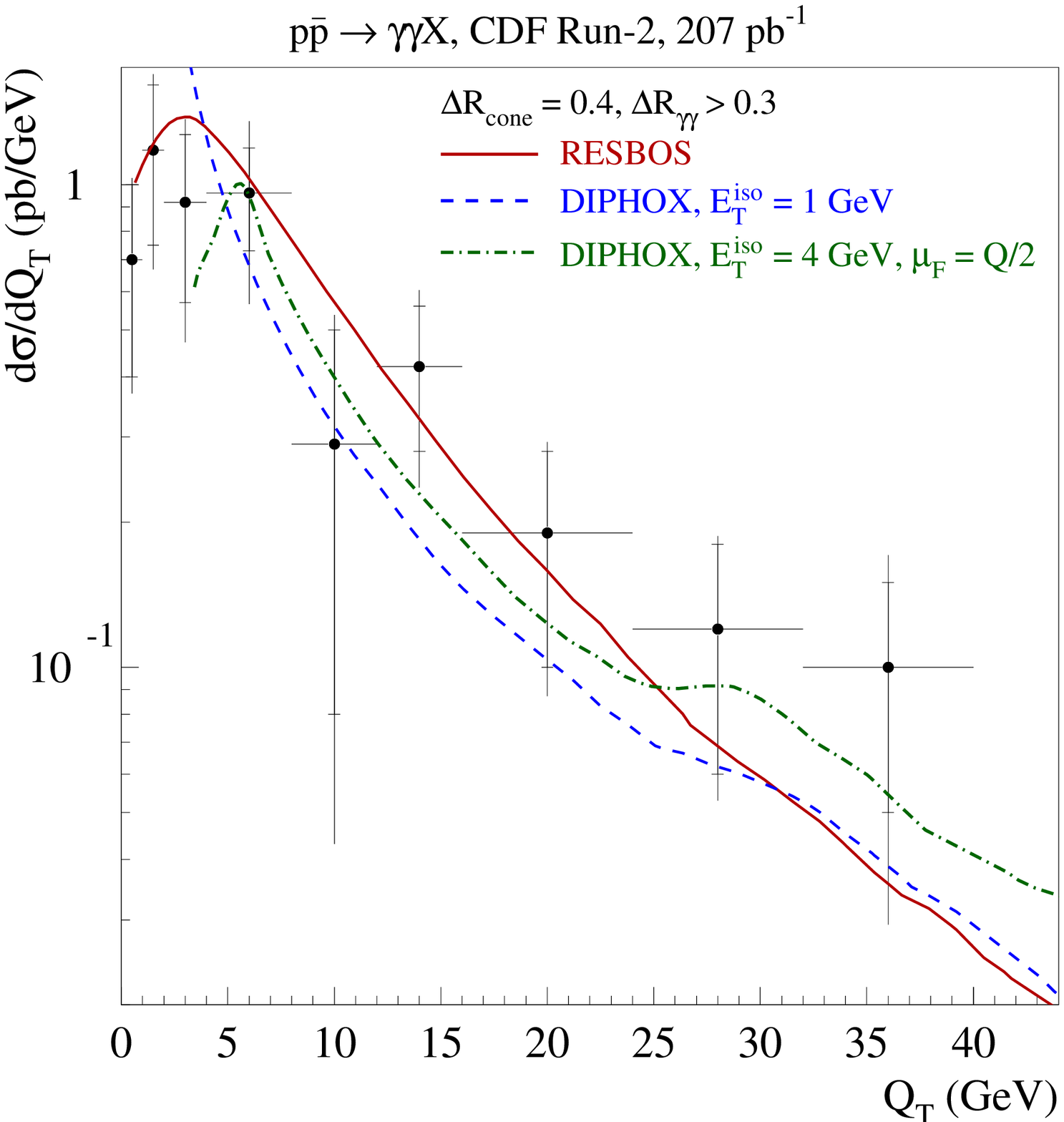}\\
(a)\hspace{2in}(b)\end{center}

\caption{(a) Invariant mass and (b) transverse momentum distributions of diphotons.  
Data from the CDF Run-2 measurement~\cite{Acosta:2004sn} are compared to our 
calculation (RESBOS) and the DIPHOX calculation.\label{fig:QQT}}
\end{figure}

The invariant mass ($Q$) distribution is shown in Fig.~\ref{fig:QQT}a.
It exhibits a characteristic lower kinematic cutoff at 
$Q\approx2\sqrt{p_{Tmin}^{\gamma_{1}}p_{Tmin}^{\gamma_{2}}} \approx 27$
GeV. Our calculation (RESBOS) agrees well with the data.  In this figure 
we also show the perturbative QCD contributions evaluated at finite order, 
represented by the DIPHOX code~\cite{Binoth:1999qq}.  Unless specified 
otherwise, the scales $\mu_{R}=\mu_{F}=Q$ are used to obtain the DIPHOX 
results presented here.  
The overall agreement between the two calculations is anticipated,
since both evaluate the inclusive rates at NLO accuracy.
The differences are due to different isolation prescriptions, resummation 
of higher-order logarithms as well as NLO contributions to the $gg$ channel 
in our calculation, and single-photon one- and two-fragmentation contributions in DIPHOX.  

The transverse momentum ($Q_{T}$) distribution of diphotons is 
shown in Fig.~\ref{fig:QQT}b.  The finite-order calculation, represented here 
by DIPHOX, displays an unphysical logarithmic singularity as 
$Q_T \rightarrow 0$.  In our work, the initial-state small-$Q_{T}$
singularities are resummed in the CSS formalism, resulting in a reasonable
overall shape of the cross section at any $Q_{T}$.  
The fragmentation contributions exhibit a double-logarithmic singularity when $Q_T$
approaches $E_T^{iso}$ from below~\cite{Binoth:1999qq}, as it is evident in the DIPHOX
$Q_T$ distribution for $E_T^{iso} = 4$ GeV. No such singularity is present in our 
$Q_T$ distribution, which instead has a mild discontinuity at the point $Q_T =
E_T^{iso}$ where we switch from the quasi-experimental to smooth-cone isolation.

For the same value $E_{T}^{iso}=1$ GeV, our distributions and those of 
DIPHOX agree well at large $Q_{T}$, as a result of our smooth matching of the
resummed cross section to the NLO cross section. In the
two highest-$Q_{T}$ bins, the CDF central values lie above the two
theory predictions. While the observed excess of events in this {}``shoulder''
region is not significant compared to the present experimental errors,
it has been discussed as a possible indication of enhanced fragmentation
contributions in the Tevatron data~\cite{Acosta:2004sn,Binoth:2000zt}.

\begin{figure}
\begin{center}\includegraphics[%
  width=0.5\textwidth,
  keepaspectratio]{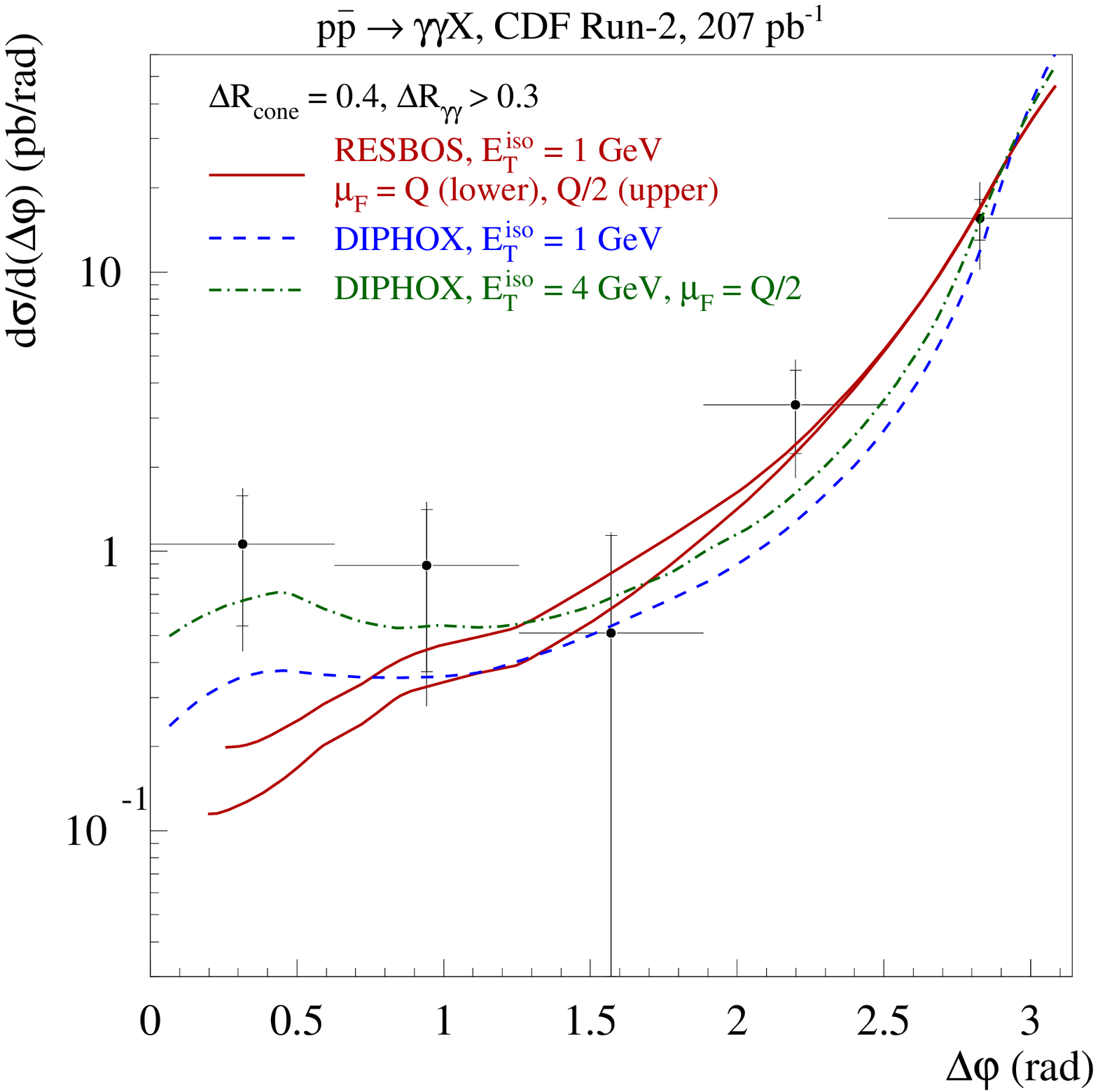}~~\includegraphics[%
  width=0.5\textwidth,
  keepaspectratio]{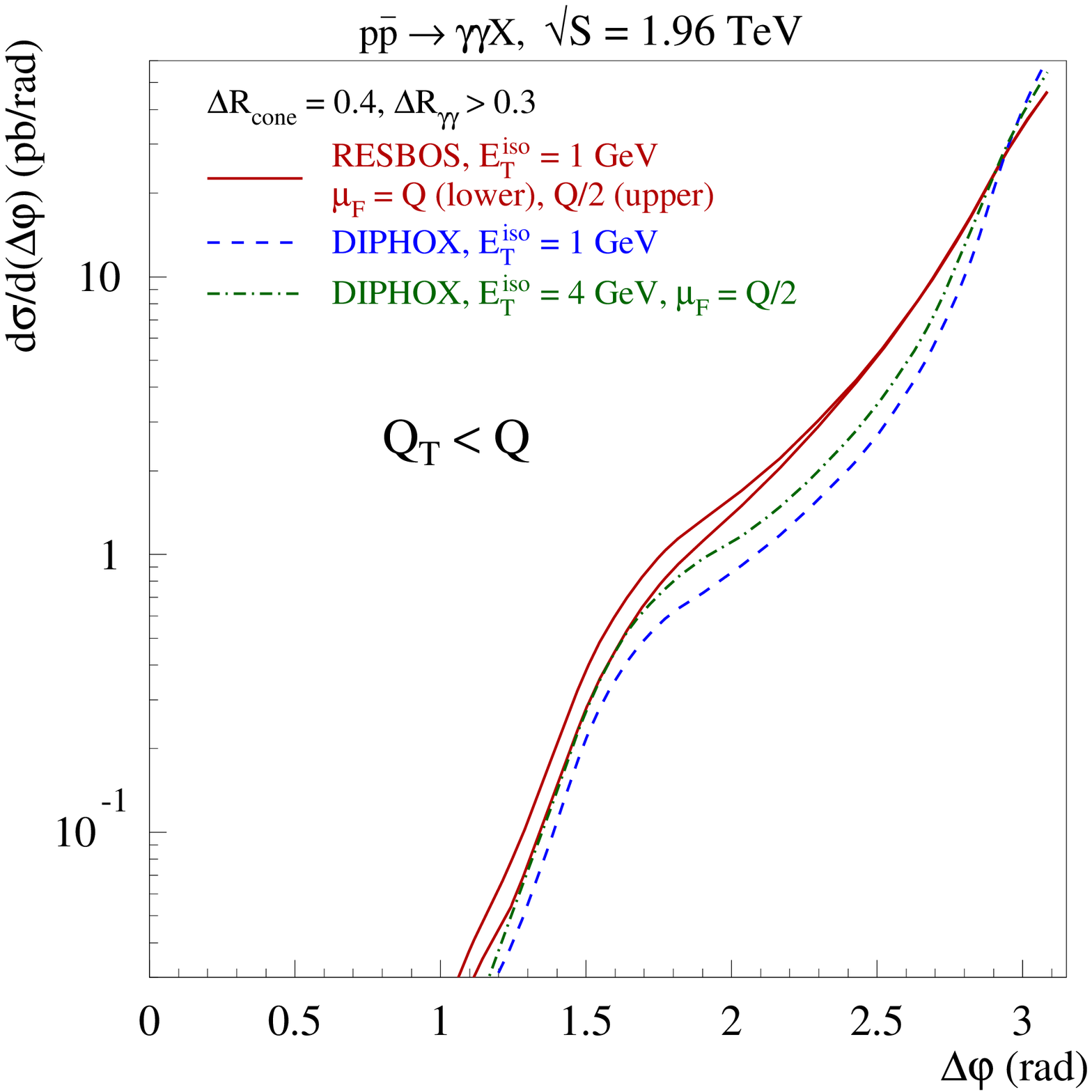}\\
(a)\hspace{2in}(b)\end{center}

\caption{(a) Distribution over the azimuthal separation $\Delta\varphi$ between
the two photons.  Data from the CDF Run-2 measurement~\cite{Acosta:2004sn} are 
compared with our calculation (RESBOS) and the DIPHOX calculation for different 
isolation parameters;
(b) same as (a), with an additional cut $Q_{T}<Q$ on the diphoton momentum.
Our cross sections are evaluated with the factorization scales
$\mu_{F}=Q$ (lower curve) and 0.5~ $Q$ (upper curve) in the finite-order contribution.
\label{fig:Dphi}}
\end{figure}

The parameters in DIPHOX can be adjusted to bring its results into agreement
with the data in the shoulder region (cf. the dash-dot curves in Figs.~\ref{fig:QQT}b
and~\ref{fig:Dphi}a).  The cross section in that region is enhanced if a smaller 
factorization scale is used, and if the isolation energy $E_{T}^{iso}$ is increased. 
The dash-dot curves in Figs.~\ref{fig:QQT}b and~\ref{fig:Dphi}a 
are obtained with $\mu_F = 0.5~Q$ and 
$E_{T}^{iso} =4$~GeV, compared to the nominal value of $E_{T}^{iso} = 1$~GeV in the 
CDF publication. In the shoulder region, the 
increase in $E_{T}^{iso}$ to $4$~GeV strongly enhances the DIPHOX cross section 
to the value shown in the CDF publication.  
The magnitude of the one-fragmentation cross section associated with 
$D_{\gamma}(z)$ is increased on average by
400\% when $E_{T}^{iso}$ is increased from 1 to 4 GeV. 

Our calculations show that most of the shoulder events populate 
a limited volume of phase space characterized
by $\Delta\varphi\lesssim1$ rad, $Q<27$ GeV, and $Q_{T}\gtrsim25$
GeV. The location of the shoulder in the $Q_T$ distribution is 
sensitive to the value of the cut on the minimum transverse 
momentum, $p_{T}^{\gamma}$, of the individual photons, moving to 
larger $Q_T$ if these cuts are raised.  It has also been 
noted~\cite{Binoth:2000zt} that non-zero values of $p_{T}^{\gamma_{1}}$
and $p_{T}^{\gamma_{2}}$ disallow contributions with small $Q_T$ if 
the azimuthal angle separation between the two photons is small, 
$\Delta\varphi<\pi/2$.
The excess of the experimental rate over our prediction in the 
region $\Delta\varphi<0.6$ radian (cf.~Fig.~\ref{fig:Dphi}a) contributes 
the bulk of the excess seen in the shoulder in the $Q_{T}$ distribution 
in Fig.~\ref{fig:QQT}b.  We note, in addition, that the excess at small 
$\Delta \varphi$ and large $Q_T$ is characterized by $Q_T \gtrsim Q$.  

From a theoretical point of view, when 
$Q_{T}>Q$, as in the shoulder region, the calculation must be 
organized in a different way~\cite{Berger:1998ev,Berger:2001wr} in order
to resum contributions arising from the fragmentation of partons into a 
pair of photons with small invariant mass. In 
addition, a small azimuthal separation $\Delta\varphi$
often implies that the photons are produced at polar angles $\theta_{*}\approx0$
or $\pi$ in the Collins-Soper diphoton rest frame \cite{Collins:1977iv}.
The matrix element 
for the Born scattering process $q\bar{q}\rightarrow\gamma\gamma$
diverges as $\left|\cos\theta_{*}\right|\rightarrow1$. 
Large QCD corrections are known to exist when $\left|\cos\theta_{*}\right|\sim1$
at any order of the strong coupling strength $\alpha_{s}$.  Radiation
of additional partons at higher orders regularizes the singularity
of the quark propagator, yet the enhancement of the cross section
is still felt at large $\left|\cos\theta_{*}\right|$. At small
$Q_{T}$, the $\left|\cos\theta_{*}\right|\approx1$ contributions
are excluded by the cuts $p_{T}^{\gamma}>14$ (13) GeV on the transverse 
momenta of the individual photons.
If, however, the diphoton system is boosted
in the transverse direction ($Q_{T}>Q$), contributions with $\left|\cos\theta_{*}\right|\approx1$
and substantial rapidity separation $\left|y_{\gamma_{1}}-y_{\gamma_{2}}\right|>0.3$
are allowed in the event sample. 

Adequate treatment of the light $\gamma\gamma$ pairs and large-$\left|\cos\theta_{*}\right|$
contributions is missing in both our calculation and DIPHOX. The presence of 
higher-order contributions is reflected in the sensitivity of the 
DIPHOX prediction at small $\Delta\varphi$ to variations of $E_{T}^{iso}$, 
factorization scales, and the angular separation $\Delta R_{\gamma\gamma}$
between the photons~\cite{Binoth:2000zt}. 
In view of the theoretical uncertainties in the calculation of the
fragmentation contributions, and the likely presence of other types 
of radiative corrections, we suggest that more theoretical and
experimental effort is needed to firmly establish the origin of 
the excess rate in the CDF data at large $Q_T$ and small $\Delta \varphi$, 
whether from single-photon fragmentation as implemented in DIPHOX 
or/and other types of enhanced scattering contributions.

The theoretical ambiguities arise in a small part of phase space, where 
the cross section is also small.  Our theoretical treatment is most reliable 
in the region in which $Q_{T}<Q$.  When the $Q_T < Q$ selection is made, the 
contributions from $\Delta\varphi<\pi/2$  are
efficiently suppressed, and dependence on tunable isolation parameters
and factorization scales is reduced (cf. Fig.~\ref{fig:Dphi}b).   The
fixed-order predictions agree well between our calculation and DIPHOX, while
our resummmed cross section also provides an accurate description
of the rate at small values of $Q_{T}$.  After the  
selection $Q_T < Q$, we expect that the large $Q_T$ shoulder will disappear 
in the experimental $Q_T$  distribution.

\begin{figure}
\begin{center}\includegraphics[%
  width=0.5\textwidth,
  keepaspectratio]{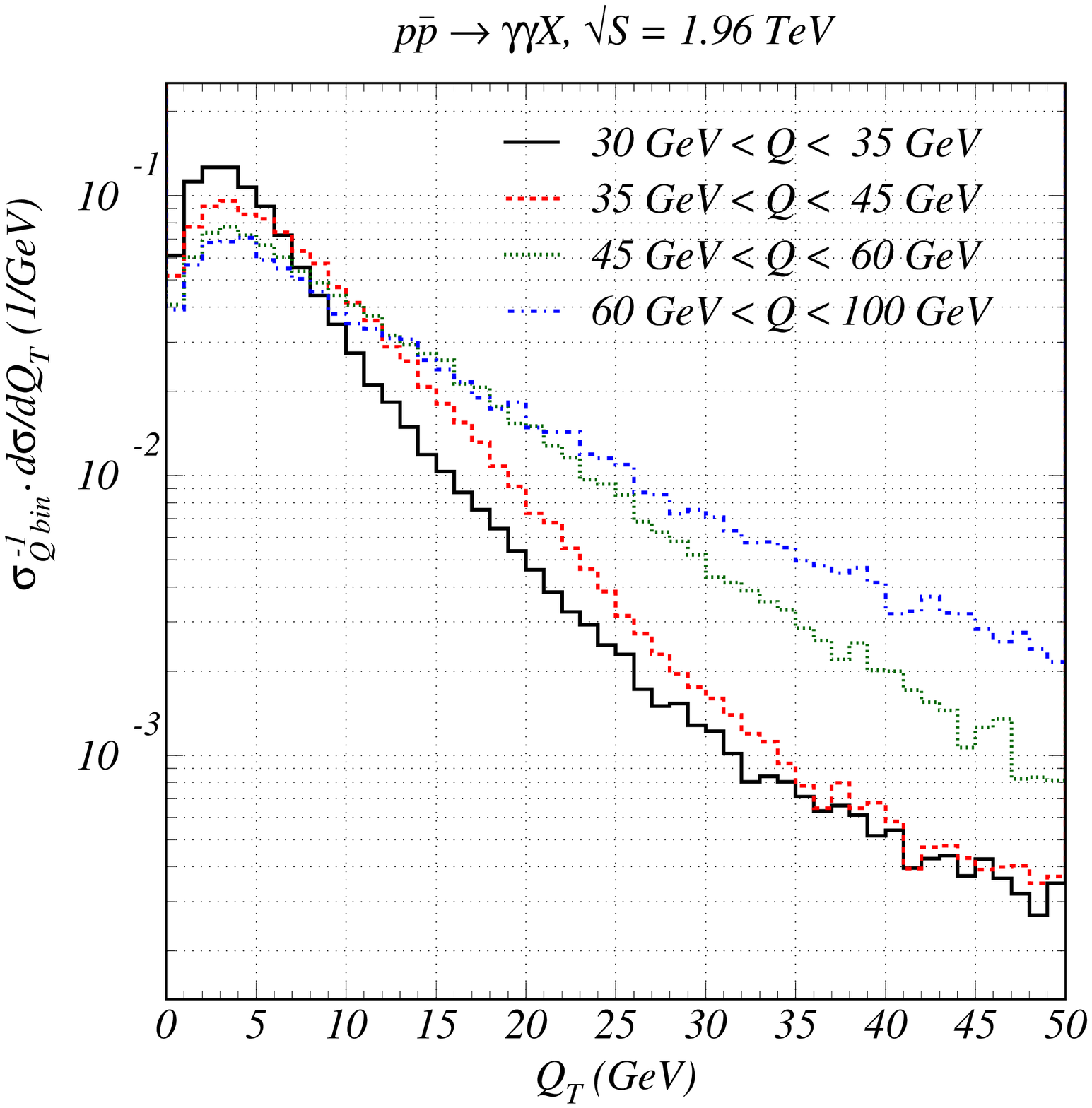}~~\includegraphics[%
  width=0.5\textwidth,
  keepaspectratio]{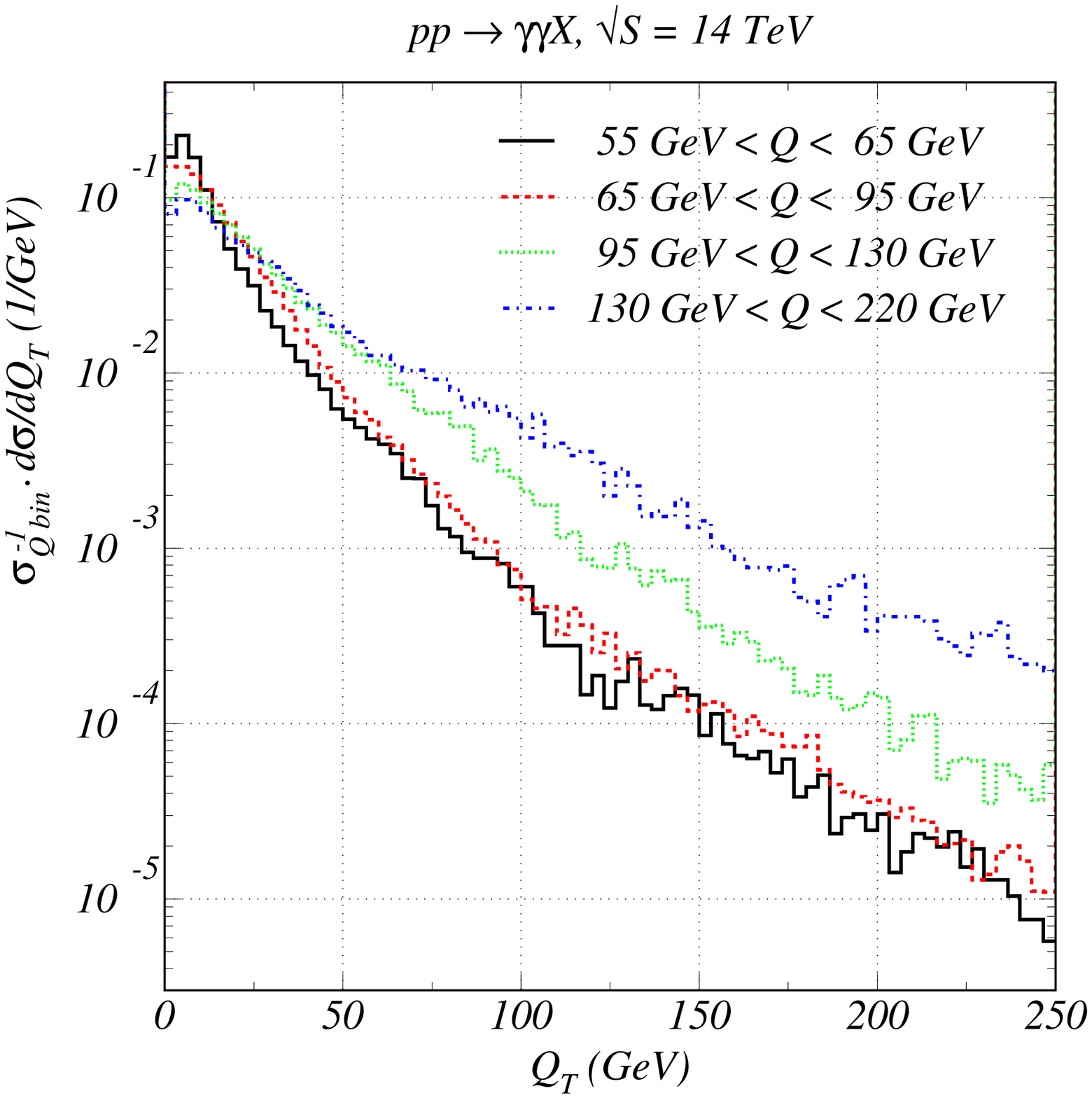}\\
(a)\hspace{2in}(b)\end{center}

\caption{Resummed transverse momentum distributions of photon pairs in 
various $\gamma \gamma$ invariant mass ($Q$) bins at (a) the Tevatron 
and (b) the LHC.
The curves are calculated with the cuts specified in the text.   
The cross sections are normalized to the total cross section in each bin of 
$Q$.  We note that our predictions are most reliable in the region $Q_T < Q$, 
but we plot the curves over the full range of $Q_T$, using the procedure 
described in the text to switch from the resummed to the 
finite-order perturbative results for $Q_T > Q$. }
\label{Fig:QTMassBins}

\end{figure}

An important prediction of the resummation formalism is a logarithmic dependence 
on the diphoton invariant mass $Q$.  In Fig.~\ref{Fig:QTMassBins}a, we show the 
resummed transverse momentum distributions for various intervals of $Q$.  The 
$Q_T$ distribution is predicted to 
broaden with increasing $Q$.  The average values of $Q_T$ are 
$\langle Q_T \rangle =$ (6.5, 8.1, 10.7, and 12.6) GeV for invariant masses in the intervals 
(30-35, 35-45, 45-60, and 60-100) GeV, respectively.  To compute these averages, 
we integrate over 
the range $Q_T = 0$ to $200$~GeV.  We urge the CDF and D0 collaborations to 
verify this predicted broadening with $Q$. 

{\em {Results for the LHC.}}  To obtain predictions for $p p$ collisions 
at $\sqrt S = 14$~TeV, we employ the following cuts on the kinematics of 
the individual photons.   For each photon, we 
require transverse momentum~$p_{T}^{\gamma}>25$~GeV and 
rapidity~$|y_{\gamma}|<2.5$.
We impose a somewhat looser isolation restriction than for the Tevatron study,
requiring less than $E_T^{iso} = 10$ GeV of extra  
transverse energy inside a cone with $\Delta R = 0.7$ around each photon.

Figure~\ref{Fig:QTMassBins}b shows the resummed transverse momentum 
distributions for various selections of diphoton invariant mass at the LHC. 
The plot shows the broadening of the $Q_T$ distribution with increasing 
mass: in the ranges (55-65, 65-95, 95-130, and 130-250) GeV the 
values of $\langle Q_T \rangle$ are (14, 17, 25, and 33) GeV.  At the LHC, we integrate 
from $Q_T = 0$ to $250$~GeV to obtain the averages.  
For the mass range appropriate in the search for a Standard Model Higgs boson, e.g.,  
$115$ to $130$~GeV, the diphoton background that 
we consider in this paper has $\langle Q_T \rangle \sim 27$~GeV, to be compared with the 
expectation for the signal of $\sim 40$~GeV~\cite{Berger:2002ut}.  The harder 
transverse momentum 
distribution for the signal arises because their is more soft gluon radiation 
in the dominant gluon-fusion Higgs boson production process~\cite{Berger:2002ut}.  
Additional predictions for the LHC are presented in Ref.~\cite{DiphotonLongPRD}.

{\em Summary.} We present a new QCD calculation of the transverse 
momentum distribution of photon pairs produced at hadron colliders, 
including all-orders resummation of initial-state soft-gluon radiation 
valid at next-to-next-to-leading logarithmic accuracy.  This calculation 
is most appropriate for values of $\gamma \gamma$ transverse momentum 
$Q_T$ not in excess of the $\gamma \gamma$ invariant mass $Q$.  Resummation 
changes both the shape and normalization of the $Q_T$ distribution, with 
respect to a finite-order calculation, in the range of values of $Q_T$ 
where the cross section is largest.  Comparison of our results with 
data from the Fermilab Tevatron shows good agreement, and we offer 
suggestions for a more differential analysis of the Tevatron data. We 
also include predictions for the Large Hadron Collider.
  
Our calculation accounts for the effects of soft gluon radiation on 
transverse momentum distributions through all orders of $\alpha_{s}$. 
The NLO calculation with inclusion
of single-photon fragmentation \cite{Binoth:1999qq} is another important
approach to $\gamma\gamma$ production. However, theoretical uncertainties  
are present in the rate of fragmentation contributions associated with 
the kinematic approximations and tunable parameters in the 
quasi-experimental isolation condition.  For $Q_{T}>Q$
($\Delta\varphi<\pi/2$), new types of higher-order contributions are 
expected to enhance the rate above our predictions.  The interpretation of
the region of small $\Delta\varphi$ remains ambiguous, as several
distinct processes may contribute to the enhanced rate. This interesting 
region warrants further theoretical investigation.  With the contributions
from the $Q_{T}>Q$ region removed, our calculation describes   
the leading contributions in the $q\bar{q} + qg$ and $gg$ diphoton 
production channels at NNLL accuracy.

\section*{Acknowledgments}

We acknowledge helpful discussions with R.~Blair, J.~Proudfoot, J.~Huston, 
J.-P. Guillet, and Y.~Liu.  
Work at Argonne is supported in part by the U.~S. Department of Energy, 
Division of High Energy Physics, Contract
W-31-109-ENG-38. The work of C.-P.~Y. is supported by the U.~S. 
National Science Foundation under award PHY-0244919.  We 
acknowledge the use of \textit{Jazz}, a 350-node computing
cluster operated by the Mathematics and Computer Science Division
at ANL as part of its Laboratory Computing Resource Center.

\end{document}